\documentclass[aps,prb,twocolumn, showpacs]{revtex4}
\usepackage{amsmath}
\usepackage{amssymb}
\usepackage{bm}
\usepackage{epsfig}
\usepackage{graphicx}
\usepackage{color}

\def\be{\begin{equation}}
\def\ee{\end{equation}}
\def\bea{\begin{eqnarray}}
\def\eea{\end{eqnarray}}

\begin{document}

\title{Spin dynamics of electrons and holes in InGaAs/GaAs quantum wells at milliKelvin temperatures}
\author{L.~V.~Fokina$^{1}$, I.~A. Yugova$^{2}$, D.~R. Yakovlev$^{1,3}$,  M.~M.~Glazov$^{3}$, I.~A.~Akimov$^{1,3}$, A.~Greilich$^{1}$, D.
Reuter$^4$, A.~D. Wieck$^4$, and M. Bayer$^1$} \affiliation{$^1$
Experimentelle Physik II, Technische Universit\"at Dortmund,
D-44221 Dortmund, Germany} \affiliation{$^2$ Department of Solid
State Physics, Physical Faculty of St.Petersburg State University,
198504 St. Petersburg, Russia} \affiliation{$^3$ A.~F. Ioffe
Physico-Technical Institute, Russian Academy of Sciences, 194021
St.Petersburg, Russia} \affiliation{$^4$ Angewandte
Festk\"orperphysik, Ruhr-Universit\"at Bochum, D-44780 Bochum,
Germany}

\begin{abstract}
The carrier spin dynamics in a \emph{n}-doped (In,Ga)As/GaAs quantum
well has been studied by time-resolved Faraday rotation and
ellipticity techniques in the temperature range down to 430
milliKelvin. These techniques give data with very different spectral
dependencies, from which nonetheless consistent information on the
spin dynamics can be obtained, in agreement with theoretical
predictions. The mechanisms of long-lived spin coherence generation
are discussed for the cases of trion and exciton resonant
excitation. We demonstrate that carrier localization leads to a
saturation of spin relaxation times at 45~ns for electrons below
4.5~K and at 2~ns for holes below 2.3~K. The underlying spin
relaxation mechanisms are discussed.
\end{abstract}
\pacs{78.67.Hc,78.47.-p,71.35.-y }
\date{\today}

\maketitle

\section{Introduction}\label{sec:intro}
The spin physics of semiconductor heterostructures attracts
considerable attention nowadays due to the emerging fields of
semiconductor spintronics, quantum computation and quantum
information~\cite{Spintronics,Spinbook,Qbits}. Understanding the
basic mechanisms providing spin orientation, spin relaxation and
spin decoherence of electrons and holes as well as manifestation of
these mechanisms in various experimental conditions, e.g. external
magnetic fields, lattice temperatures, etc., is of great importance
in this respect. One of the evident goals is to optimize material
properties and heterostructure design to achieve the longest
possible spin relaxation time so that sufficient room is left for
implementing protocols for spin manipulation and read-out.

Carrier localization quenches the particle orbital motion and is one
of the pathways to suppress efficient spin-relaxation mechanisms
related to the spin-orbit interaction. It has been shown that the
spin relaxation of electrons localized on donors in bulk GaAs can
exceed 100~ns~\cite{Kikkawa,Dzhioev}. Also in (In,Ga)As/GaAs quantum
dots the electron spin coherence time can reach 3~$\mu$s, while the
hole spin relaxation time may exceed tens of
nanoseconds~\cite{Greilich_Science, Crooker}. In quantum well
structures localization of the two-dimensional electrons at liquid
helium temperatures is required to demonstrate relaxation times in
the order of tens of nanoseconds~\cite{Zhukov06,bat_paper}. Under
these conditions spin relaxation times of few nanoseconds have been
reported for resident
holes~\cite{Marie99,Baylac95,Syperek,Kugler09}. Very recently a
remarkably long hole spin relaxation time of 70~ns has been measured
at 400~mK by a resonant spin amplification technique in
\emph{p}-doped GaAs/(Al,Ga)As quantum wells~\cite{Korn09}. This time
decreases to 2.5~ns with the temperature increase up to 4.5~K.
Evidently temperatures below those that can be obtained by pumping
$^4$He (about 1.5~K) are essential for understanding the carrier
spin dynamics in quantum wells. However, the available experimental
data are very limited in this range, mostly due to the demanding
efforts for performing experiments with $^3$He and the
complicated direct optical access to the sample in this case.

In this paper we apply time-resolved Faraday rotation (FR) and
ellipticity techniques to study the carrier spin dynamics in
\emph{n}-doped (In,Ga)As/GaAs quantum well. In Sec.~\ref{sec:exper}
we provide a description of these pump-probe techniques.
Experimental features and theoretical modeling of the detected spin
polarization are discussed in Sec.~\ref{sec:frell}. Two
contributions to the Faraday rotation and ellipticity signals due to
interaction of the probe beam with the trion and exciton resonances
are considered. In Sec.~\ref{sec:rsa} we discuss the resident
electron spin orientation by the circularly polarized pump. Two
mechanisms are suggested for the initialization of the experimentally observed long-lived
spin dynamics. In Sec.~\ref{sec:low} we focus on the long-lived spin
dynamics of resident electrons measured in the regime of resonant
spin amplification (RSA) at very low temperatures down to 430~mK.
The characteristic bat-like shape of the RSA signal contains
information on the hole spin dynamics, which persists remarkably
long at low temperatures. We discuss the mechanisms responsible for
electron and hole spin relaxation in the addressed temperature
ranges.

\section{Experimental details}\label{sec:exper}

\begin{figure}[hptb]
\includegraphics[width=\linewidth]{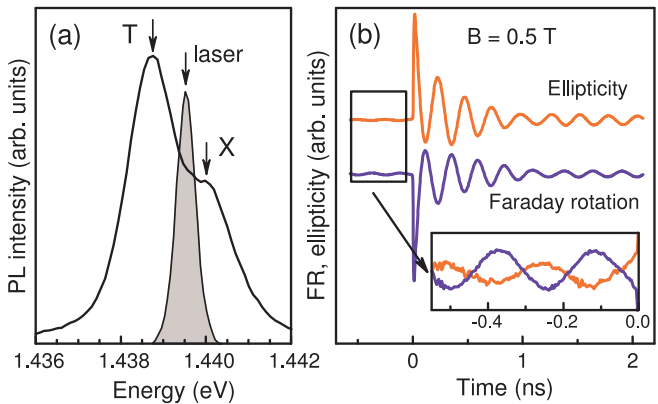}
\caption{(Color online) (a) Photoluminescence spectrum of the
(In,Ga)As/GaAs QWs studied experimentally (black line). Filled
contour represents the spectrum of the excitation laser used in the
Faraday rotation and ellipticity experiments. (b) Time-resolved Faraday rotation and
ellipticity signals measured in nondegenerate regime (pump at
1.4395~eV, probe at 1.4382~eV). Signals are vertically shifted for
clarity. $T=1.6$~K. Inset: Long-lived parts of the signals shortly
before pump pulse arrival.} \label{fig:fig1}
\end{figure}

We study a heterostructure with two coupled 8~nm thick
In$_{0.09}$Ga$_{0.91}$As/GaAs quantum wells (QWs) separated by a
thin (1.7~nm) GaAs barrier. The layer sequence was grown on an
undoped GaAs substrate with (100) orientation by molecular-beam
epitaxy. It contains a 100~nm \emph{n}-doped GaAs buffer layer
separated by a 100~nm GaAs spacer from the QWs. The doped layer
serves as source of resident electrons for the QWs. The
two-dimensional electron gas density in the QWs does not exceed
$10^{10}$~cm$^{-2}$ in absence of optical excitation. This structure
was selected on purpose because of pronounced carrier localization
at cryogenic temperatures~\cite{bat_paper}. The structure was
optically characterized measuring the photoluminescence (PL) spectra
[Fig.~\ref{fig:fig1}(a)]. In the PL spectrum two emission lines
separated by 1.4~meV are observed which we attribute to exciton (X)
and negatively charged trion (T) recombination~\cite{bat_paper}.
Their full width at half maximum of 1~meV is caused by trion and
exciton localization on alloy and QW width fluctuations.

In our pump-probe setup, we used one (degenerate regime) or two
(nondegenerate regime) synchronized Ti:Sapphire lasers, each with
pulse duration of 5 ps and pulse repetition period of 13.2~ns
corresponding to a repetition frequency of 75.6 MHz. The laser
wavelength was tuned in resonance with either the QW exciton or
trion transitions. The circular polarization of the pump beam was
modulated at 50~kHz by an elasto-optical modulator. The excitation
density was kept close to the lowest possible limit at
0.5~W/cm$^{2}$. A sensitive balanced photodiode scheme was used for
recording the Faraday rotation and ellipticity signals. The sample
was mounted in a cryostat with a split-coil superconducting magnet
which allowed us to perform experiments in the Voigt configuration
with the magnetic field $\textbf{B}\parallel \textbf{x}$ oriented in
the quantum well plane perpendicular to the light propagation
direction (coinciding with the structure growth axis, denoted as
$z$-axis). The sample temperature was varied from 0.34~K to 80~K. For
milliKelvin measurements the cryostat had a $^3$He inset, instead of
a $^4$He variable temperature inset. The sample was then in contact
with $^3$He in the inset, whose temperature could be changed between
0.3 and 100~K. Windows on the $^3$He inset allow direct optical
access to the sample. When using this milliKelvin inserts we tried
to minimize the laser power as much as possible to avoid overheating
of the sample.

\section{Detection of Faraday rotation and ellipticity signals}\label{sec:frell}

Figure~\ref{fig:fig1}(b) shows Faraday rotation and ellipticity
signals obtained in the nondegenerate regime at $B=0.5$~T. The pump
energy was approximately tuned to the exciton resonance and the
signal was probed at a different energy close to the trion
resonance. The Faraday rotation and ellipticity signals are similar
to each other and contain a fast decaying part with a decay time of
360~ps, which is the typical recombination time for excitons and
trions in these QWs. Therefore we assign this component to the spin
dynamics of photogenerated carriers~\cite{Gerlovin, Zhukov}.
Additionally a long-lived component can be seen at times exceeding
1~ns, which is even found at delays of about 13~ns, i.e. shortly
before the next pump pulse arrival, as shown in the inset of
Fig.~\ref{fig:fig1}(b). The long-lived signal is caused by
generation of spin coherence for the resident electrons and its
decay is solely controlled by the electron spin dephasing time,
which may be as long as 55~ns in the studied
structure~\cite{bat_paper}.

The scenarios of spin coherence generation for the resident
electrons in quantum wells have been considered in
Ref.~\onlinecite{Zhukov}. As a rule, the spin orientation of
resident carriers involves either direct trion formation or trion
formation from a spin oriented exciton generated by the pump. The
trion formation is accompanied by capture of a resident electron
with defined spin orientation, which results in polarization of the
resident electron ensemble. Another mechanism can become involved
for resonant exciton excitation: due to the exchange scattering
between a resident electron and the electron from a photocreated
exciton the resident carriers can become spin polarized. We
demonstrate below in Sec.~\ref{sec:rsa} that these two mechanisms
can be equally important under resonant exciton excitation.

\begin{figure}[hptb]
\includegraphics[width=\linewidth]{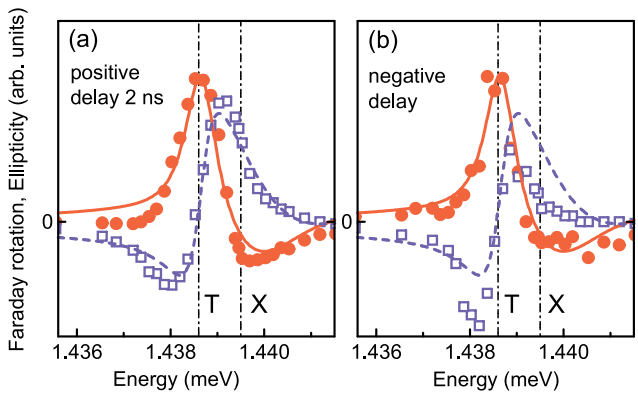}
\caption{(Color online) Dependencies of long-lived signal amplitudes
of Faraday rotation (open squares and dashed lines) and ellipticity
(closed circles and solid lines) on spectral position of probe.
$B=0.5$~T and $T=1.6$~K. Experimental data are shown in panel (a)
for a positive pump-probe delay of 2~ns and in panel (b) for a small
negative delay. Pump at 1.4395~eV. Lines in both panels are fits by
Eqs.~(\ref{signalsT}) and (\ref{signalsX}) with the following
parameters: $\hbar \Gamma_\mathrm{T} =0.65$~meV, $\hbar
\bar{\Gamma}_\mathrm{X} =2.5$~meV, $\hbar
\beta_\mathrm{X}/\alpha_\mathrm{T}=1.56$~meV, $\hbar
\omega_\mathrm{T}=1.4386$~eV, $\hbar \omega_\mathrm{X}=1.4395$~eV.}
\label{fig:fig2}
\end{figure}

Now we turn to the spectral dependencies of the long-lived
oscillations in the Faraday rotation and ellipticity  signals, which
are shown in Fig.~\ref{fig:fig2}. The experimental data measured at
relatively large positive pump-probe delay of 2~ns and a small
negative delay (shortly before pump pule arrival) are given in
panels (a) and (b), respectively. The Faraday rotation and
ellipticity signals obviously demonstrate quite different spectral
behaviors. The maximum of the ellipticity signal and the zero of the
FR signal are close to the trion resonance energy. Also, there is an
extra feature in the ellipticity signal around the exciton
resonance.

The Faraday rotation, $\mathcal F$, and ellipticity, $\mathcal E$,
of the probe pulse are proportional to the imaginary and real parts
of the difference of quantum well transmission coefficients for
$\sigma^+$ and $\sigma^-$ polarizations, $t_{\pm}$, respectively.
Taking into account that for the relatively thin quantum wells
studied here $t_{\pm} = 1+r_{\pm}$ where  $r_\pm(\omega)$ are the
corresponding quantum well reflection
coefficients\cite{Zhukov,FR_and_Ell}
\begin{equation}
\label{signals}
\mathcal E + \mathrm i \ \mathcal F \propto r_+(\omega) - r_-(\omega).
\end{equation}
Equation~\eqref{signals} is valid provided that $r_\pm(\omega)\ll 1$
which is the case for the studied system. The difference between
$r_+$ and $r_-$ arises from the pump-induced spin polarization of
the resident carriers. The quantum well reflection coefficients
contain contributions from the exciton (X) and trion (T) resonances:
\begin{equation}
\label{reflect} r_{\pm}(\omega) = \sum_{i=\rm T, X} \frac{\mathrm i
\Gamma_{0,i}^{\pm}}{\omega_{i}^\pm -\omega -\mathrm i
(\Gamma_{0,i}^\pm + \Gamma_i^\pm)}.
\end{equation}
Here $\omega_i^\pm$ are the resonance frequencies of exciton ($i=$X)
and trion ($i=$T) for the corresponding circular polarizations,
$\Gamma_{0,i}^\pm$ and $\Gamma_i^\pm$ are the radiative and
non-radiative dampings, respectively (see Section 3.1.1 in
Ref.~\onlinecite{ivchenko}).

\begin{figure}[hptb]
\includegraphics[width=\linewidth]{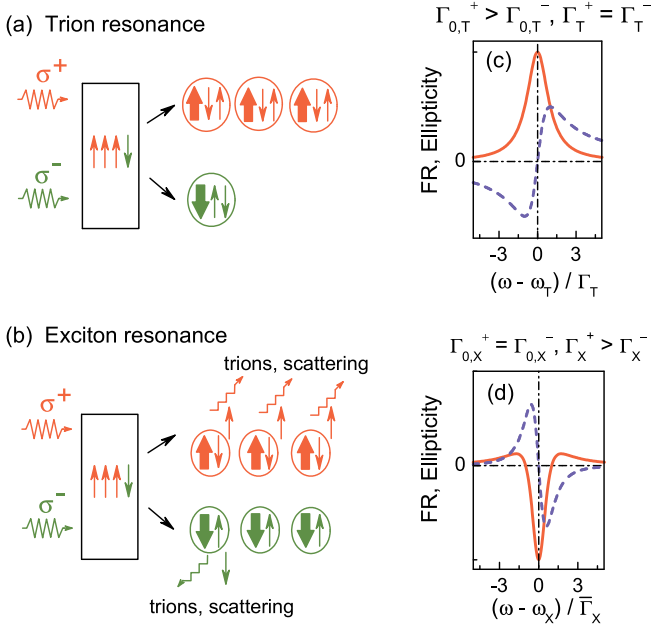}
\caption{(Color online) Schematic illustration of the Faraday
rotation and ellipticity signal formation for probing the trion
[panels (a) and (c)] and exciton [panels (b) and (d)] resonances.
Thin arrows show electron spins, thick arrows show $z$ component of
a photocreated hole spin. The imbalance of spin-up and spin-down
resident electrons results in different interaction efficiencies of
the $\sigma^+$ and $\sigma^-$ probe components with the resident
carriers.  In case of trion resonance detection the trion oscillator
strength for a given polarization is directly proportional to the
number of electrons with corresponding spin $z$ component, i.e. if
$N_+>N_-$ the $\sigma^+$ component of the linearly polarized probe
is reflected more efficiently compared with the $\sigma^-$ component
because more trions can be formed then. In case of exciton resonance
detection the damping rates for excitons created by the
$\sigma^+$ and $\sigma^-$ components of the probe are different. For
$N_+>N_-$ the non-radiative broadening of $\sigma^+$ created
excitons is higher since they participate more efficiently in the
formation of trions and in the exchange electron-exciton scattering
processes and the reflectivity in this polarization is smaller.}
\label{fig:scheme1}
\end{figure}

When the spin polarized resident electrons are probed by linearly
polarized light different physical processes are responsible for the
Faraday rotation and ellipticity signals, depending on whether the
probe is resonant with either the trion or the exciton resonance.
This is explained schematically in Fig.~\ref{fig:scheme1}. The
linearly polarized probe pulse can be decomposed into two circularly
polarized components, one of which interacts with the spin-polarized
resident electrons more efficiently compared with the other.

For the trion resonance shown in Figs.~\ref{fig:scheme1}(a) and
\ref{fig:scheme1}(c) the main modulation contribution is caused by
the trion radiative broadening $\Gamma_{0,\mathrm{T}}^\pm$ (trion oscillator strength). The
trion oscillator strength in the $\sigma^\pm$ polarizations is
proportional to the number of electrons with $z$ spin component $\pm
1/2$, $N_{\pm}$. Hence, $\Gamma_{0,\mathrm{T}}^\pm=\alpha_{\mathrm
T} \Gamma_{0,\mathrm X} N_\pm$, where $\Gamma_{0,\rm X}$ is the
exciton radiative broadening calculated by neglecting electron spin
polarization, and $\alpha_{\mathrm T}$ is a constant related with
the effective trion area, see Ref.~\onlinecite{ast00}. As a result,
the difference of reflection coefficients in Eq.~\eqref{signals}
takes the form
\begin{equation}
\label{signalsT} \mathcal E + \mathrm i \ \mathcal F \propto
\frac{\mathrm i \alpha_{\rm T}\Gamma_{0,\rm X}(N_+-N_-)}{\omega_{\rm
T} -\omega -\mathrm i \Gamma_{\rm T}},
\end{equation}
where we exploited the fact that the non-radiative broadening
$\Gamma_\mathrm{T}$ exceeds by far the radiative one,
$\Gamma_\mathrm{{0,T}}$. The corresponding shapes of the FR and ellipticity
signals are shown in  Fig.~\ref{fig:scheme1}(c). The FR is an odd
function of the detuning of the probe energy from the trion
resonance, while the ellipticity is an even function.

For the exciton resonance shown in Figs.~\ref{fig:scheme1}(b) and
\ref{fig:scheme1}(d) the situation is different compared to the
trion resonance. At low temperatures and in presence of spin
polarized resident electrons the dominant modulation effect results
from the spin dependent non-radiative damping of the excitons:
$\Gamma_{\rm X}^\pm = \bar{\Gamma}_{\rm X} + \beta_{\rm X} N_\pm$.
Here $\bar{\Gamma}_{\rm X}$  is the exciton non-radiative
broadening, which does not depend on the exciton spin orientation.
The spin-dependent part of the exciton broadening, $\beta_{\rm X}
N_\pm$, is caused by two processes: the exchange electron-exciton
scattering and the trion formation by the photogenerated exciton and
a spin-polarized resident electron, see Fig.~\ref{fig:scheme1}(b). Here
$\beta_{\rm X}$ is a constant characterizing the efficiency of these
spin-dependent mechanisms~\cite{ast00,Zhukov}. As a result the
exciton contributions to the ellipticity and FR signals has the form
\begin{equation}
\label{signalsX} \mathcal E + \mathrm i \ \mathcal F \propto
-2\frac{\Gamma_{0,\rm X} \beta_{\rm X}(N_+-N_-)}{(\omega_{\rm X}
-\omega -\mathrm i \bar{\Gamma}_{\rm X})^2},
\end{equation}
where we also neglected $\Gamma_{0, \rm X}$ compared with
$\bar\Gamma_{\rm X}$. The corresponding exciton contributions to the
FR and ellipticity signals are plotted in Fig.~\ref{fig:scheme1}(d). The
signs of the signals are inverted as compared with the trion case in
Fig.~\ref{fig:scheme1}(c). This is because for a given circular
polarization the presence of spin polarized electrons decreases the
reflectivity of an exciton, while the reflectivity of a trion is
enhanced~\cite{ast00}. The ellipticity signal has its minimum at the
exciton resonance frequency and changes its sign at the wings, while
the FR signal is an odd function of the detuning between exciton
resonance and probe optical frequency.

In Figs.~\ref{fig:fig2}(a) and \ref{fig:fig2}(b) the measured
dispersions of the FR and ellipticity signals are compared with the
theoretical model. The signals are the superpositions of the exciton
and trion contributions, as clearly seen at the higher energy side
of the ellipticity signal measured for positive delays where it changes sign,
Fig.~\ref{fig:fig2}(a). A similar trend is also observed for
negative delays,  Fig.~\ref{fig:fig2}(b), however the exciton
contribution is less pronounced there.

We attribute
the differences in the spectral behavior of the measured signals at
positive and negative delays to the fact, that there are two
subensembles of resident electrons, see e.g.
Ref.~\onlinecite{Zhukov}. The electrons with stronger localization
demonstrate longer spin relaxation time. They make a major
contribution to the pump-probe signal at negative delays and
modulate the signal at the trion frequency more efficiently as
compared with the less localized carriers. The latter have shorter
spin relaxation times so that their contribution is more pronounced
in the signals at positive delays. This conclusion is supported by
the observation, that two Larmor frequencies which lie quite close
to each other are detected for the long-lived signal, see below.

\section{Spin coherence initialization and resonant spin amplification}\label{sec:rsa}

The observation of electron spin polarization before the next pump
pulse arrival indicates that the spin relaxation time is comparable
or longer than the laser repetition period of 13.2~ns and that spin
polarization may accumulate from pulse to pulse. The resonant spin
amplification technique (RSA) allows us to extract the long spin
relaxation times with high accuracy~\cite{Kikkawa}. In this section
we discuss results of pump-probe Faraday rotation RSA experiments.
We checked also that the ellipticity RSA data give the same spin
relaxation times as the FR signals.

\begin{figure}[hptb]
\includegraphics[width=\linewidth]{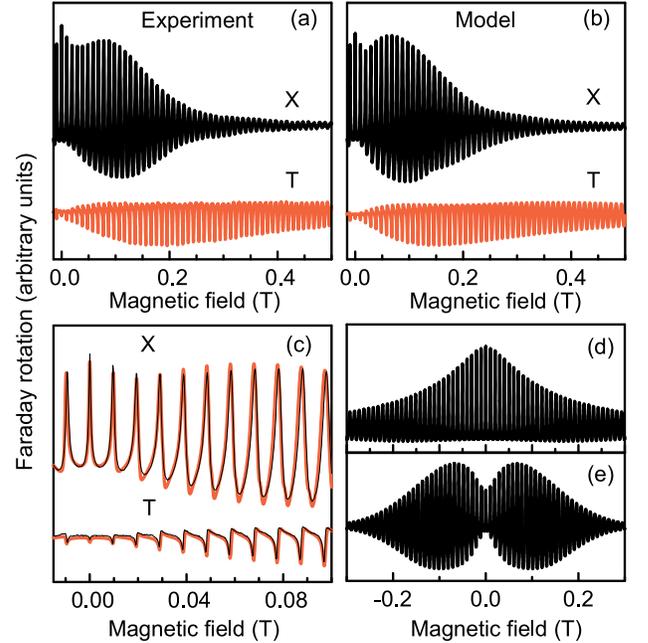}
\caption{(Color online) (a) and (c) RSA signals measured in
degenerate pump/probe regime at the trion and exciton resonances at
$T=1.8$~K.  Black curves in panel (c) are RSA experimental data and
thick red (grey) curves are fits by Eq.~(\ref{eq:sz}). (b) Modeled
RSA signals for the trion and exciton resonances. The two components
in the theoretical RSA signal for exciton resonance due to spin
orientation by exciton or trion absorption are shown separately in
panels (d) and (e), respectively. Calculation parameters are:
$T^e_s=45$~ns and $\tau_r=120$~ps. For RSA on trion resonance:
$|g_e|=0.555$, $\Delta g_e=0.002$,  $T^h_s=2.5$~ns. For RSA on
exciton resonance: (i) spin orientation by excitons $|g_e|=0.535$,
and $\Delta g_e=0.003$; (ii) spin orientation by trions:
$|g_e|=0.550$, $\Delta g_e=0.008$, $T^h_s=1.8$~ns.} \label{fig:fig3}
\end{figure}

Figure~\ref{fig:fig3} shows RSA signals measured for degenerate
pump/probe conditions at the trion and exciton resonance energies.
The peaks in the RSA signals correspond to the spin precession
frequencies which are commensurable with the laser repetition
frequency. From the peak width the electron spin relaxation time
$T_s^e$, which in this case corresponds to the spin dephasing time
$T^*_{2,e}$ of the electron spin ensemble precessing about the
magnetic field, can be evaluated~\cite{Kikkawa}.

The spin amplification signals measured on the exciton and trion
resonance qualitatively differ from each other, see
Fig.~\ref{fig:fig3}(a). They deviate also from the typical RSA shape
with decreasing peak amplitude and increasing peak width with
increasing magnetic field, see Ref.~\onlinecite{Kikkawa} and
Fig.~\ref{fig:fig3}(d). For the trion resonance the signal amplitude
is strongly suppressed at zero magnetic field and the amplitude
increases with growing field strength. For the exciton resonance the
signal has a complicated shape, which results from a combination of
the typical RSA signal from Fig.~\ref{fig:fig3}(d) and the trion
bat-like signal shown in Fig.~\ref{fig:fig3}(e) (see below).

\begin{figure}[h]
\includegraphics[width=0.95\linewidth]{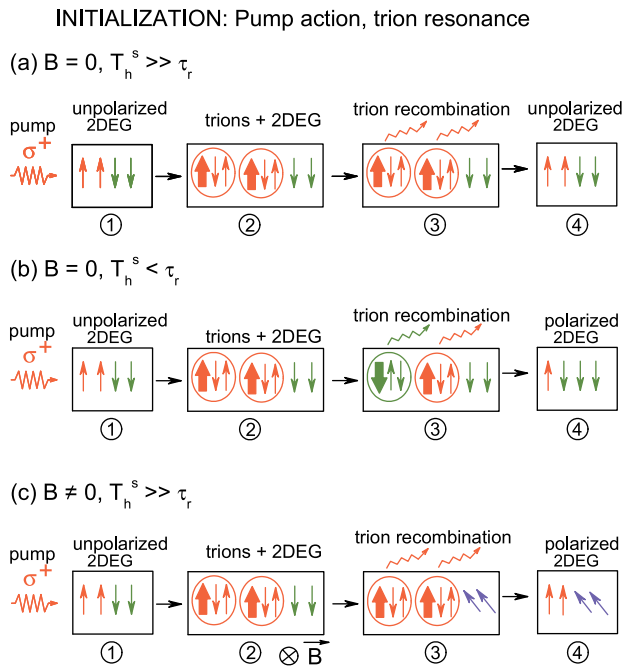}
\caption{(Color online)  Spin orientation of resident electrons
denoted as 2DEG (two-dimensional electron gas) for trion resonant
excitation. The following four stages are shown: (1) unpolarized
2DEG before excitation; (2) result of action of a
$\sigma^+$-polarized pump pulse, part of the resident electrons are
bound to trions; (3) trions and resident electrons shortly before
trion recombination; (4) 2DEG after trion recombination. Panels (a)
and (b) show the situation for zero external magnetic field. (a)
$T_s^h\gg\tau_r$, hole spin flip is absent and resident electrons
stay unpolarized after trion recombination. (b) $T_s^h< \tau_r$,
hole spin relaxes before trion decay and resident electrons become
spin polarized. (c) Non-zero external magnetic field. Even in the
absence of hole spin relaxation the resident electrons become
polarized due to electron spin precession about magnetic field
during trion lifetime.} \label{fig:scheme2tr}
\end{figure}

The strong difference between the signals measured at the exciton
and trion energies suggests that different mechanisms are involved
in spin coherence generation. The strong suppression of the RSA
signal at $B=0$ for trion excitation and its bat-like shape serves
as direct evidence of the long spin relaxation time of the hole
involved in the trion~\cite{bat_paper}. Indeed, in absence of the
magnetic field the long-lived electron spin coherence at trion
excitation can be generated only as much as the hole spin in trion
flips, compare Figs.~\ref{fig:scheme2tr}(a) and
\ref{fig:scheme2tr}(b). If the hole spin relaxation time $T_s^h$
exceeds by far the trion recombination time, $\tau_r$, the resident
electron left behind after trion recombination has the same spin as
before and no long-living spin coherence for the resident electrons
is generated, Fig.~\ref{fig:scheme2tr}(a). With increase of the
magnetic field the resident carrier spins and the spins of the
electron returning from the trion are no longer parallel to each
other and the RSA signal increases, see Fig.~\ref{fig:scheme2tr}(c).

\begin{figure}[h]
\includegraphics[width=0.95\linewidth]{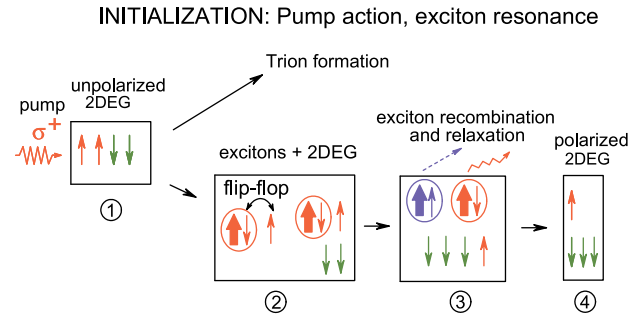}
\caption{(Color online) Spin orientation of resident electrons at
exciton resonant excitation. Two possible scenarios are depicted:
trion formation from the photocreated exciton and flip-flop
scattering of an exciton with a resident electron. In this case the
exciton is transformed into ta dark state and the resident electrons
become polarized after the dark state has decayed (shown by blue
dashed arrow).} \label{fig:scheme2ex}
\end{figure}

Interestingly, for exciton resonant excitation (see
Fig.~\ref{fig:fig3}) the signal is not suppressed at $B=0$, which
shows that electron spin coherence is efficiently excited 
even in the absence of electron spin rotation. The corresponding mechanism of resident
electron spin coherence generation can be related with the spin
flip-flop exchange scattering between a resident electron and a
photoexcited electron in the exciton. In such a case the hole
spin-flip is not needed: right after the electron flip-flop scattering an
uncompensated resident electron spin appears
(Fig.~\ref{fig:scheme2ex}). This exchange scattering process may be
the dominant channel for spin coherence generation at weak magnetic
fields, while with increasing magnetic field the spin coherence
generation via the trion state becomes more efficient. This results
in the increase of the RSA amplitude for fields $B>50$~mT, see
Fig.~\ref{fig:fig3}(a).

The bat-like trion signal shown in Fig.~\ref{fig:fig3}(a) can be
modeled with good quantitative agreement, as seen from the
comparison of the black and grey lines in Fig.~\ref{fig:fig3}(c), in
the frame of the approach suggested in  Ref.~\onlinecite{bat_paper}.
An infinite sequence of pump pulses with repetition period $T_R$
creates an electron spin polarization periodic in time
\cite{Shabaev, bat_paper}. The electron spin polarization,
$\widetilde{s}_{z}$, shortly before pump pulse arrival is described
by~\cite{bat_paper,Note1}:
\begin{widetext}
\begin{eqnarray}
       \widetilde{s}_z&=&\frac{1}{2}\left [\frac{K}{1+ue^{-2T_R/T^e_s}-(1+u)\cos(\omega_eT_R)e^{-T_R/T^e_s}-K} \right ],\label{eq:sz}\\ \nonumber
K &\equiv& v\left [u\alpha e^{-T_R/T^e_s}- \alpha \cos(\omega_eT_R)-\beta\sin(\omega_eT_R)\right]e^{-T_R/T^e_s}.
\end{eqnarray}
\end{widetext}
Here $v \equiv \sin^2(\Theta/2)/2$, $u \equiv \cos(\Theta/2)$,
$\Theta$ is pulse area, $\alpha \equiv 1+Re(1/(\tau_r\Omega))$,
$\beta \equiv Im(1/(\tau_r\Omega))$, $\Omega
=1/T^e_s-1/\tau_T+i\omega_e$, $T_R$ is laser repetition period,
$\omega_e=g_{e}\mu_B B/\hbar$ is the electron spin precession
frequency, $\tau_r$ is the trion recombination time,
$\tau_T=T_s^h\tau_r/(T_s^h+\tau_r)$ is the trion spin lifetime,
$T^h_s$ is the hole spin relaxation time, and $T^e_s$ is the
electron spin relaxation time.

The expression for a typical RSA contour which describes the
contribution due to spin orientation by electron-exciton scattering
can be found, for example, in the online materials to
Ref.~[\onlinecite{Greilich_Science}] as Eqs.~(B4) and (B5). It is
described by Eq.~(\ref{eq:sz}) with
$K=v\left[ue^{-T_R/T^e_s}-\cos(\omega_eT_R)\right]e^{-T_R/T^e_s}.$
An example of such signal is shown in Fig.~\ref{fig:fig3}(d).

In order to fit the complicated shape of the RSA signal measured at
the exciton resonance we used a sum of contributions due to spin
orientation by trion and by exciton. The grey line fit for the
exciton in Fig.~\ref{fig:fig3}(c), results from two signals shown
separately in Figs.~\ref{fig:fig3}(d) and \ref{fig:fig3}(e). These
two contributions have not only different shapes but also slightly
different electron $g$-factors and spreads of $g$-factors, as given
in the figure caption. The variation of $g$-factors can be traced to
its energy dependence~\cite{Yugova_factor}. The weaker localized
carriers, which have a stronger impact on the exciton resonance,
have a smaller $g$-factor and a smaller dispersion.

It is worth mentioning that for exciton resonant excitation higher
energy trions can be formed as compared to the case of trion
resonant excitation. These trions can be weaker localized so that
they demonstrate short hole spin relaxation times, resulting in
efficient spin coherence generation for resident electrons at $B=0$.
The contribution of such a channel can be described by the
conventional RSA shape shown in Fig.~\ref{fig:fig3}(d).
However, in the studied sample the role of the exciton under these experimental conditions dominated over the possible trion contribution. There are two arguments for this conclusion: (i) the spectral dependence of amplitude of the RSA signal with usual shape [Fig.~\ref{fig:fig3}(d)] has maximum at exciton pumping and (ii) this amplitude increases with increasing of pump power and disappears with decrease of pump power. It allows us to conclude that the main contribution to the RSA signal with shape shown in Fig.~\ref{fig:fig3}(d) is due to electron spins polarized by electron-exciton scattering or due to fast spin relaxation of hole in the resonantly excited exciton.

\section{Low temperature spin dynamics}\label{sec:low}

\begin{figure}[hptb]
\includegraphics[width=0.9\linewidth]{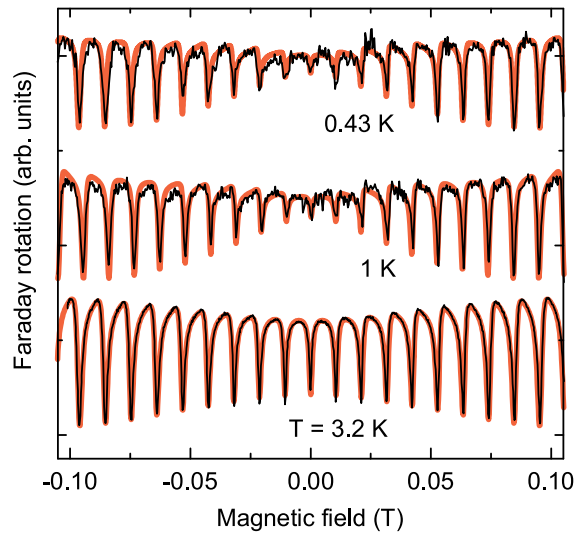}
\caption{(Color online) RSA signals measured by degenerate
pump-probe Faraday rotation at the trion resonance for various
temperatures. Black curves are experimental data and thick red
(grey) curves are fits by Eq.~(\ref{eq:sz}). Calculation parameters
are: $|g_e|=0.555$, $T^e_s=45$~ns, for $T = 0.43$~K: $T^h_s=2$~ns,
 $\tau_r=200$~ps; for $T = 1$~K: $T^h_s=2$~ns,
 $\tau_r=200$~ps; for $T = 3.2$~K: $T^h_s=0.6$~ns,
 $\tau_r=120$~ps.}
\label{fig:fig4}
\end{figure}

We turn now to the evolution of the spin dynamics of electrons and
holes and in particular to the mechanisms providing carrier spin
relaxation at extremely low temperatures down to 430 mK. We focus on
the Faraday rotation RSA signals measured at the trion resonance,
where the hole spin dynamics is most pronounced.
Figure~\ref{fig:fig4} shows RSA signals measured at the trion
resonance energy for various temperatures. We fit the experimental
RSA spectra from Fig.~\ref{fig:fig4} by Eq.~(\ref{eq:sz}) and get
very good agreement in all cases. The $g$-factors and spin
relaxation times determined in that way are given in the figure
caption.

\begin{figure}[hptb]
\includegraphics[width=0.9\linewidth]{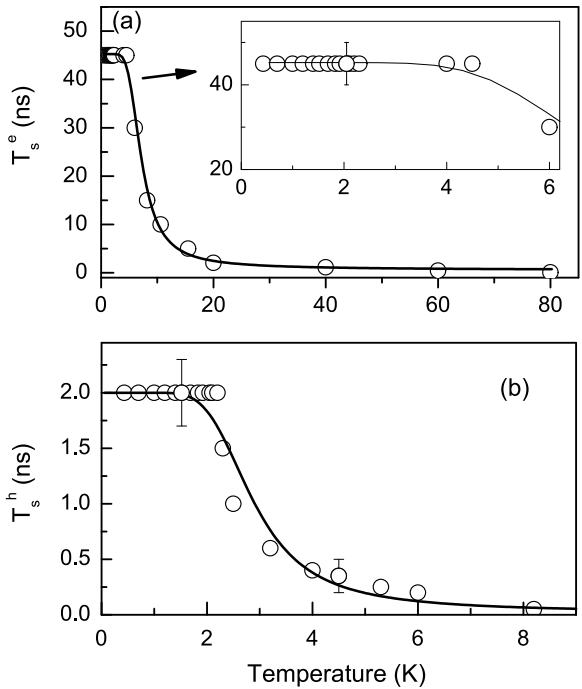}
\caption{Temperature dependencies of electron (a) and hole (b) spin
relaxation times in an (In,Ga)As/GaAs QW. Symbols are experimental
data and lines are fits with Eq.~(\ref{Ts}) using the following
parameters: $T^e_0=45$~ns, $T^e_{exc}=0.5$~ns and $\Delta E_e=3$~meV
for electrons in panel (a), and $T^h_0=2$~ns, $T^h_{exc}=10$~ps and
$\Delta E_h=1.4$~meV for holes in panel (b).} \label{fig:fig5}
\end{figure}

The electron and hole spin relaxation times measured at different
temperatures are collected in Fig.~\ref{fig:fig5}. The data for
temperatures below 15~K were determined from RSA spectra and for
higher temperatures, at which the RSA signals vanish, we fit the
decay of the FR signals  at positive delays. The data in the
temperature range 0.4-10~K have been measured using the $^3$He
insert, and in the temperature range 1.6-80~K in a $^4$He insert.
The two temperature ranges were chosen to have overlap, in order to
confirm consistency of the data.

The temperature dependencies of spin relaxation times for electrons
and holes are qualitatively similar. The times almost do not change
at very low temperatures, which shows that the spin dynamics in
these regimes are controlled by temperature independent mechanisms.
At elevated temperatures the times drop by more than an order of
magnitude. As can be seen from the solid lines in
Fig.~\ref{fig:fig5} these behaviors can be well fitted by the
following function describing thermal activation from the ground
state with a long relaxation time to an excited state with a shorter
relaxation time:
\begin{equation}
\frac{1}{T_s^{e(h)}}=\frac{1}{T_0^{e(h)}}+\frac{1}{T_{exc}^{e(h)}}\exp{\left(-\frac{\Delta
E_{e(h)}}{k_BT}\right)}. \label{Ts}
\end{equation}
Here $T_0^{e(h)}$ are the spin relaxation times in the electron (hole) ground states, $T_{exc}^{e(h)}$ are constants
characterizing the transitions between the ground and excited states
which depend on the electron-phonon interaction, $\Delta E_{e(h)}$
are the characteristic activation energies, and $k_B$ is the
Boltzmann constant. One should note that this approach is rather
simplified and gives physically feasible values of the fitting
parameters for a relatively narrow temperature range. Therefore, it
should be valid for holes in a temperature range not exceeding 8~K,
while for electrons, for which experimental data have been recorded
up to 80~K, additional mechanisms may cause the temperature
dependencies of $T_{exc}^{e}$ and $\Delta E_{e}$ so that more
elaborated approach may be required.

The electron spin relaxation time, $T_s^e$, is constant in the
temperature range from 0.43 to 4.5~K at an extremely large value of
45~ns, see insert in Fig.~\ref{fig:fig5}(a). At these temperatures
the resident electrons are localized and their spin relaxation is
provided by the hyperfine interaction with the nuclei spins, which
is almost temperature independent.  At temperatures above
5~K the resident electrons are thermally activated and the
Dyakonov-Perel relaxation mechanism~\cite{Spinbook} which is very
efficient for free electrons starts to act. As a result the
relaxation time drops to 1.1~ns at 40~K and further down to 110~ps
at 80~K. Such a behavior is typical for \emph{n}-doped QWs with a
diluted concentration of the resident
electrons~\cite{Zhukov06,Gerlovin,bat_paper}.

Let us turn now to the temperature dependence of the hole spin
relaxation time, $T_s^h$, in Fig.~\ref{fig:fig5}(b). Below 2.2~K
$T_s^h$ saturates at a value of 2~ns. At the moment it is not clear
to us what relaxation mechanism is controlling the hole spin
dynamics in the range from 0.43-2.2~K. Most probably it is due to
the admixture of the light-hole to the heavy-hole states, enhancing
strongly the possibility for spin-flip scattering. 

The hyperfine interaction with the nuclei is significantly weaker
for the holes than for the electrons. This has been confirmed by the
recent report on ultralong hole spin relaxation with about 70~ns
relaxation time, measured in the range from 0.4-1.2~K for a
\emph{p}-doped 4-nm GaAs/Al$_{0.3}$Ga$_{0.7}$As QW~\cite{Korn09}.
Our experimental situation differs from the one in \emph{p}-doped
samples with resident holes, as in \emph{n}-doped samples we detect
the spin dynamics of photogenerated holes bound in negatively
charged trions. However, in the trion ground state, which is a
singlet state, the two electrons have antiparallel spin orientations
and flip-flop electron-hole process are not possible without
exciting the trion complex into a triplet state, which requires an
energy similar to the trion binding energy of 1.4~meV. In fact this
energy is in very good agreement with the activation energy $\Delta
E_h=1.4$~meV, which has been obtained from fitting the experimental
data in Fig.~\ref{fig:fig5}(b) by Eq.~(\ref{Ts}). Therefore we
suggest that the strong decrease of the hole spin relaxation time at
temperatures above 2.2~K is either due to trion thermal
dissociation, which excites the hole into the continuum of free
states with a strong spin-orbit interaction leading to a fast spin
relaxation, or due to flip-flop process involving
the trion triplet state.

In conclusion, the carrier spin dynamics in an \emph{n}-doped
(In,Ga)As/GaAs QW have been studied by the resonant spin
amplification technique at very low lattice temperatures down to
0.43~K. Carrier localization leads to a saturation of the spin
relaxation times at 45~ns for electrons below 4.5~K and at 2~ns for
holes below 2.3~K. Also the spectral dependencies of the Faraday
rotation and ellipticity signals have been studied experimentally
around the trion and exciton resonances. The mechanisms responsible
for spin polarization of the resident electrons under resonant
pumping into the trion and exciton resonances have been discussed.

We greatly acknowledge an expert assistance of D.~Fr\"ohlich for the experiments performed at milliKelvin temperatures. The work has been supported by the Deutsche Forschungsgemeinschaft
and the EU Seventh Framework Programme (Grant No. 237252,
Spin-optronics). I.A.Y. thanks the Ministry of Education and Science
of the Russian Federation (Grants 2.1.1.1812 and 02.740.11.0214).
M.M.G. acknowledges support by the President grant for young
scientists, the "Dynasty" Foundation -- ICFPM and the Russian
Foundation for Basic Research.

\end{document}